\documentclass[11pt]{article}

\usepackage{a4wide}
\usepackage{graphicx}
\usepackage{color}
\usepackage[hypertex]{hyperref}

\title{A Stochastic Evolutionary Growth Model for Social Networks}

\author{Trevor Fenner, Mark Levene, George Loizou and George Roussos \\
School of Computer Science and Information Systems \\
Birkbeck College, University of London \\
London WC1E 7HX, U.K. \\ \{trevor,mark,george,gr\}@dcs.bbk.ac.uk}

\date{}

\begin{document}

\maketitle

\newtheorem{theorem}{Theorem}[section]
\newtheorem{corollary}[theorem]{Corollary}
\newtheorem{lemma}[theorem]{Lemma}
\newtheorem{proposition}[theorem]{Proposition}
\newtheorem{definition}{Definition}[section]
\newtheorem{algorithm}{Algorithm}
\newtheorem{example}{Example}[section]

\begin{abstract}

We present a stochastic model for a social network, where new actors may join the network, existing
actors may become inactive and, at a later stage, reactivate themselves. Our model captures the
evolution of the network, assuming that actors attain new relations or become active according to
the preferential attachment rule. We derive the mean-field equations for this stochastic model and
show that, asymptotically, the distribution of actors obeys a power-law distribution. In particular,
the model applies to social networks such as wireless local area networks, where users connect to
access-points, and peer-to-peer networks where users connect to each other. As a proof of concept,
we demonstrate the validity of our model empirically by analysing a public log containing traces
from a wireless network at Dartmouth College over a period of three years. Analysing the data
processed according to our model, we demonstrate that the distribution of user accesses is
asymptotically a power-law distribution.

\end{abstract}

\section{Introduction}

We present a stochastic model for a social network \cite{SCOT00}, where new actors may join the
network, existing actors may become inactive and, at a later stage, may reactivate themselves. Our
model captures the evolution of the network, assuming that actors attain new relations or become
active according to the preferential attachment rule. The concept of {\em preferential attachment},
originating from \cite{PRIC76}, has become a common theme in stochastic models of networks
\cite{ALBE01,NEWM03}. This behaviour often results in the ``rich get richer'' phenomenon, for
example, where new relations to existing actors are formed in proportion to the number of relations
those actors currently have.

\smallskip

The model presented incorporates the novel aspect of differentiating between active and inactive
actors, and allowing actors' status to change between active and inactive over time. This type of
network dynamics is especially relevant to situations where actors may connect/disconnet or
login/logout from the network, in particular, when network registration is needed as a prior
condition to the first time an actor connects to the network. The network models proposed so far
either assume that all actors are active, or that when actors leave the network they do not rejoin
it \cite{AMAR00}.

\medskip

By deriving the mean-field equations for this model of a social network, we obtain the result that,
asymptotically, the distribution of actors obeys a power law. Power-law distributions taking the
form
\begin{displaymath}
f(i) = C \ i^{- \phi},
\end{displaymath}
where $C$ and $\phi$ are positive constants, are abundant in nature \cite{SCHR91}. The constant
$\phi$ is called the {\em exponent} of the distribution. Examples of such distributions are:
{\em Zipf's law}, which states that the relative frequency of a word in a text is inversely
proportional to its rank, {\em Pareto's law}, which states that the number of people whose
personal income is above a certain level follows a power-law distribution with an exponent
between 1.5 and 2 (Pareto's law is also known as the {\em 80:20 law}, stating that about 20\%
of the population earn 80\% of the income) and {\em Lotka's law}, which states that the number
of authors publishing a prescribed number of papers is inversely proportional to the square of
the number of publications.

Recently, several researchers have detected power-law distributions in the topology of several
networks such as the World-Wide-Web \cite{BROD00}, e-mail networks \cite{EBEL02}, collaboration
networks \cite{GROSS02b,FENN05} and peer-to-peer networks \cite{RIPE02}.

\medskip

There are several examples of networks that can be modelled within our formalism. One example is
that of a wireless network \cite{KOTZ05b}, where mobile users having, e.g. a laptop, PDA or mobile
phone, connect to access points within a defined region (e.g. campus, building or airport). In this
case the actors are the users and the relations are between users and access points. The user is
active during a connection and otherwise inactive. Another example, is that of a peer-to-peer
network \cite{ORAM01}, where users (referred to as peers) connect to other peers in order to
exchange information. Peer-to-peer networks are of prime importance to the future of the internet,
as networks such as Bittorrent \cite{POUW05}, Kazaa \cite{LIAN06} and Skype \cite{GUHA06} are
becoming increasingly popular and thus account for a sizeable amount of all internet traffic.

\medskip

Our stochastic model is based on the transfer of {\em balls} (representing actors) between {\em
urns} (representing actor states), where we distinguish between active balls in, regular, {\em
unstarred urns} and inactive balls in {\em starred urns}. The relationships of a particular actor
are represented as {\em pins} attached to the corresponding ball.

We note that our urn model is an extension of the stochastic model proposed by Simon in his
visionary paper published in 1955 \cite{SIMO55}, which was couched in terms of word frequencies in a
text. Previously, in \cite{FENN05}, we considered an alternative extension of Simon's model by
adding a preferential mechanism for discarding balls from urns resulting in an exponential cutoff in
the power-law distribution.

\smallskip

In the model we present here, at each step of the stochastic process, with probability $p$, two
events may happen: either a new active ball is added to the first unstarred urn with probability
$r$, or with probability $1-r$ an inactive ball is selected preferentially from a starred urn and is
activated by moving it to the corresponding unstarred urn. Alternatively, with probability $1-p$, an
active ball is selected preferentially from an unstarred urn and then two further events may happen:
it is either moved along to the next unstarred urn with probability $q$, or with probability $1-q$
the selected ball becomes inactive by moving it to the corresponding starred urn. We assume that a
ball in the $i$th urn has $i$ pins attached to it (which represents an actor having $i$ relations).
Our main result is that the steady-state distribution of this model is an asymptotic power law, and,
moreover, as a proof of concept we demonstrate the validity of our model by analysing data from a
real wireless network.

\medskip

The rest of the paper is organised as follows. In Section~\ref{sec:urn} we present an urn transfer
model allowing balls to be active or inactive by moving from starred urns to unstarred urns and vice
versa. We then derive in Section~\ref{sec:power-law} the steady-state distribution of the model,
which, as stated earlier, follows an asymptotic power-law distribution. In Section~\ref{sec:param}
we show how we can fit the parameters of the model to data, and in Section~\ref{sec:p2p} we
demonstrate how our model can provide an explanation of the empirical distributions found in
wireless networks. Finally, in Section~\ref{sec:concluding} we give our concluding remarks.

\section{An Urn Transfer Model}
\label{sec:urn}

We now present an {\em urn transfer model} for a stochastic process that emulates the situation
where balls (which might represent actors) become inactive with a small probability, and can later
become active again with some probability. We assume that a ball in the $i$th urn has $i$ pins
attached to it (which might represent the actors' relations). The model is an extension of our
previous model of exponential cutoff \cite{FENN02}, where balls are discarded with a small
probability.

\medskip

We assume a countable number of ({\em unstarred}) urns, $urn_1, urn_2, urn_3, \ldots \ $ and
correspondingly a countable number of {\em starred} urns $urn^*_1, urn^*_2, urn^*_3, \ldots \ $,
where the former contains active balls and the latter contain the inactive balls. Initially all of
the urns are empty except $urn_1$, which has one ball in it. Let $F_i(k)$ and $F^*_i(k)$ be the
number of balls in $urn_i$ and $urn^*_i$, respectively, at stage $k$ of the stochastic process, so
$F_1(1) = 1$, all other $F_i(1) = 0$ and all $F^*_i(1) = 0$. Then, at stage $k+1$ of the stochastic
process, where $k \ge 1$, one of two events may occur:

\renewcommand{\labelenumi}{(\roman{enumi})}
\begin{enumerate}
\item with probability $p$, $0 < p < 1$, one of two events may happen:

\begin{enumerate}

\item with probability $r$, $0 < r \le 1$, a new ball (with one pin attached to it) is
inserted into $urn_1$, or

\item with probability $1-r$, a starred urn is selected, with $urn^*_i$ being
selected with probability proportional to $i F^*_i(k)$, the number of pins it contains,
and a ball is chosen from the selected urn, $urn^*_i$, and transferred to $urn_i$ (this is
equivalent to making the ball active).

\end{enumerate}

\item with probability $1 - p$ an urn is selected, with $urn_i$ being selected with
probability proportional to $i F_i(k)$, the number of pins it contains, and a ball is
chosen from the selected urn, $urn_i$; then,

\begin{enumerate}
\item with probability $q$, $0 < q \le 1$, the chosen ball is transferred to
$urn_{i+1}$, (this is equivalent to attaching an additional pin to the ball chosen
from $urn_i$), or

\item with probability $1 - q$ the ball chosen is transferred to $urn^*_i$ (this is
equivalent to making the ball inactive).
\end{enumerate}
\end{enumerate}
\smallskip

We note that we could modify the initial conditions so that, for example, $urn_1$ and $urn^*_1$
initially contained $\delta, \delta^* > 1$ balls, respectively, instead of $urn_1$ having just one
ball and $urn^*_1$ being empty. It can be shown, from the development of the model below, that any
change in the initial conditions will have no effect on the asymptotic distribution of the balls in
the urns as $k$ tends to infinity, provided the process does not terminate with either all of the
unstarred urns empty or all of the starred urns empty (cf. \cite{FENN02}). In the former case we
need to ensure that $p > (1-p) (1-q)$, i.e. that the number of balls going into unstarred urns is
greater than the number of balls going out of unstarred urns. In the latter case we need to ensure
that $(1-p) (1-q) > p (1-r)$, i.e. that the number of balls going into starred urns is greater than
the number of balls going out of starred urns.

\smallskip

More specifically, the probability of termination must be small, i.e.
\begin{displaymath}
\left( \frac{(1-p) (1-q)}{p} \right)^\delta < \epsilon
\end{displaymath}
and
\begin{displaymath}
\left( \frac{p (1-r)}{(1-p) (1-q)} \right)^{\delta^*} < \epsilon
\end{displaymath}
for some $\epsilon > 0$. We observe that these are the probabilities that the gambler's fortune
will {\em not} increase forever \cite{ROSS83}.

\medskip

The expected total number of balls in the unstarred urns at stage $k$ is given by
\begin{eqnarray}\label{eq:balls}
E \Big( \sum_{i=1}^k F_i(k) \Big) & = & 1 + (k - 1) \Big(p - (1 - p)
(1 - q) \Big) \nonumber \\
& = & (1-p) (2-q) + k  \Big(p - (1 - p) (1 - q) \Big),
\end{eqnarray}
and in the starred urns by
\begin{eqnarray}\label{eq:starred-balls}
E \Big( \sum_{i=1}^k F^*_i(k) \Big) & = & (k - 1) \Big( (1-p) (1-q) - p (1-r) \Big).
\end{eqnarray}
\medskip

The total number of pins attached to balls in $urn_i$ at stage $k$ is $i F_i(k)$, so the
expected total number of pins in the unstarred urns is given by
\begin{eqnarray}\label{eq:pins}
E \Big( \sum_{i=1}^k i F_i(k) \Big) & = & 1 + (k - 1) \Big(r p + (1-p) q \Big) + p (1-r)
\sum_{j=1}^{k-1} \psi_j - (1-p) (1-q) \sum_{j=1}^{k-1} \theta_j,
\end{eqnarray}
where $\psi_j$, $1 \le j \le k-1$, is the expectation of $\Psi_j'$, the number of pins attached to
the ball chosen at step (ib) of stage $j+1$ (i.e. the urn number), and $\theta_j$, $1 \le j \le
k-1$, is the expectation of $\Theta_j'$, the number of pins attached to the ball chosen at step
(iib) of stage $j+1$ (i.e. the urn number). More specifically,
\begin{equation}\label{eq:psi-define}
\psi_j = E(\Psi_j') = E \left( \frac{\sum_{i=1}^j i^2 F^*_i(j)}{\sum_{i=1}^j i F^*_i(j)} \right)
\end{equation}
and
\begin{equation}\label{eq:theta-define}
\theta_j = E(\Theta_j') = E \left( \frac{\sum_{i=1}^j i^2 F_i(j)}{\sum_{i=1}^j i F_i(j)} \right).
\end{equation}
\smallskip

The quotient of sums in the second expectation in (\ref{eq:psi-define}) (respectively in
(\ref{eq:theta-define})), which we denote by $\Psi_j$ (respectively by $\Theta_j$), is the expected
value of $\Psi_j'$ (respectively of $\Theta_j'$) given the state of the model at stage $j$.

\smallskip

Correspondingly, the expected total number of pins in the starred urns is given by
\begin{eqnarray}\label{eq:starred-pins}
E \Big( \sum_{i=1}^k i F^*_i(k) \Big) & = & (1-p) (1-q) \sum_{j=1}^{k-1} \theta_j - p
(1-r) \sum_{j=1}^{k-1} \psi_j.
\end{eqnarray}

\medskip

Since at stage $j+1$ there cannot be more than $j$ pins in the system, it follows that

\begin{displaymath}\label{eq:thetaj-bounds}
1 \le \theta_j, \psi_j \le j.
\end{displaymath}
\medskip

Now let
\begin{displaymath}\label{eq:thetaj}
\theta^{(k)} = \frac{1}{k} \sum_{j=1}^k \theta_j.
\end{displaymath}
and
\begin{displaymath}\label{eq:spij}
\psi^{(k)} = \frac{1}{k} \sum_{j=1}^k \psi_j.
\end{displaymath}
\smallskip

Since there are at least as many pins (starred pins) in the system as there are balls
(starred balls), it follows from, (\ref{eq:balls}) and (\ref{eq:pins}), and,
(\ref{eq:starred-balls}) and (\ref{eq:starred-pins}), that
\begin{equation}\label{eq:thetak-bounds}
(1-p) (1-q) - p (1-r) \le (1-p) (1-q) \theta^{(k)} - p (1-r) \psi^{(k)} \le (1-p) - p (1-r),
\end{equation}
which implies that $\theta^{(k)} - \psi^{(k)}$ is bounded. This bounded difference will suffice for
the purpose of the developments in the next section and we will denote $\theta^{(\infty)}$ by
$\theta$ and $\psi^{(\infty)}$ by $\psi$.


\section{Derivation of the Steady State Distribution}
\label{sec:power-law}

Following Simon \cite{SIMO55}, we now state the mean-field equations for the urn transfer
model. For $i > 1$ we have
\begin{equation}\label{eq:unstarred}
E_k(F_i(k+1)) = F_i(k) + \beta_k \Big(q (i-1) F_{i-1}(k) - i F_i(k) \Big) + \alpha_k
(1-r) i F^*_i(k),
\end{equation}
where $E_k(F_i(k+1))$ is the expected value of $F_i(k+1)$ given the state of the model at
stage $k$, and
\begin{equation}\label{eq:betak}
\beta_k = \frac{1 - p}{\sum_{i=1}^k \ i F_i(k)},
\end{equation}
\begin{equation}\label{eq:alphak}
\alpha_k = \frac{p}{\sum_{i=1}^k \ i F^*_i(k)}
\end{equation}
are the normalising factors.

\smallskip

Equation~\ref{eq:unstarred} gives the expected number of balls in $urn_i$ at stage $k+1$. This
is equal to the previous number of balls in $urn_i$ plus the probability of adding a ball to
$urn_i$ minus the probability of removing a ball from  $urn_i$, and finally plus the
probability of transferring a ball to $urn_i$ from $urn^*_i$.

\smallskip

The first probability is just preferentially choosing a ball from $urn_{i-1}$ and transferring it to
$urn_i$ in step (iia) of the stochastic process defined in Section~\ref{sec:urn}, the second
probability is that of preferentially choosing a ball from $urn_i$ in step (iia) of the process, and
the third probability is that of preferentially transferring a ball from $urn^*_i$ to $urn_i$ in
step (ib) of the process.

\medskip

In the boundary case, $i = 1$, we have
\begin{equation}\label{eq:initial}
E_k(F_1(k+1)) = F_1(k) + p r - \beta_k \ F_1(k) + \alpha_k (1-r) \ F_1^*(k).
\end{equation}
\smallskip

Equation~\ref{eq:initial} gives the expected number of balls in $urn_1$ at stage $k+1$, which is
equal to the previous number of balls in $urn_1$ plus the probability of inserting a new ball into
this urn in step (ia) of the stochastic process defined in Section~\ref{sec:urn} minus the
probability of preferentially choosing a ball from $urn_1$ in step (iia), and finally plus the
probability of preferentially transferring a ball to $urn_1$ from $urn^*_1$ in step (ib) of the
process.

\smallskip

For starred urns, for $i \ge 1$, corresponding to (\ref{eq:unstarred}) and
(\ref{eq:initial}), we have
\begin{equation}\label{eq:starred}
E_k(F^*_i(k+1)) = F^*_i(k) + \beta_k (1-q) i F_i(k) - \alpha_k (1- r) i F^*_i(k),
\end{equation}
where $E_k(F^*_i(k+1))$ is the expected value of $F^*_i(k+1)$ given the state of the
model at stage $k$.

\medskip

Equation~\ref{eq:starred} gives the expected number of balls in $urn^*_i$ at stage $k+1$. This is
equal to the previous number of balls in $urn^*_i$ plus the probability of preferentially
transferring a ball from $urn_i$ to $urn^*_i$ in step (iib) of the stochastic process defined in
Section~\ref{sec:urn} minus the probability of preferentially transferring a ball from $urn^*_i$ to
$urn_i$ in step (ib) of the process.

\medskip

In order to solve the equations of the model, namely (\ref{eq:unstarred}),
(\ref{eq:initial}) and (\ref{eq:starred}), we make the assumptions that, for large $k$,
the random variables $\beta_k$ and $\alpha_k$ can be approximated by constants (i.e.
non-random) values depending only on $k$. To this end we take the approximations to be

\begin{equation}\label{eq:beta-hat}
\hat{\beta}_k = \frac{1-p}{(k-1) \left(r p + (1-p) q + p (1-r) \ \psi^{(k-1)} - (1-p) (1-q) \
\theta^{(k-1)} \right)},
\end{equation}
and
\begin{equation}\label{eq:alpha-hat}
\hat{\alpha}_k = \frac{p}{(k-1) \left( (1-p) (1-q) \ \theta^{(k-1)} - p (1-r) \
\psi^{(k-1)} \right)}.
\end{equation}
\smallskip

The motivation for the above approximations is that the denominators in the definitions of
$\beta_k$ and $\alpha_k$ have been replaced by asymptotic approximations of their
expectations as given in (\ref{eq:pins}) and (\ref{eq:starred-pins}), respectively. We
note en passant that replacing $\beta_k$ by $\hat{\beta}_k$ and $\alpha_k$ by
$\hat{\alpha}_k$ results in an approximation similar to that of the ``$p_k$ model'' in
\cite{LEVE01c}, which is essentially a ``mean-field''  approach.

\smallskip

We next take the expectations of (\ref{eq:unstarred}), (\ref{eq:initial}) and
(\ref{eq:starred}). By the linearity of the expectation operator $E(\cdot)$, we obtain

\begin{equation}\label{eq:hat1}
E(F_i(k+1)) = E(F_i(k)) + \hat{\beta}_k \Big(q (i-1) E(F_{i-1}(k)) - i E(F_i(k)) \Big) +
\hat{\alpha}_k (1-r) i E(F^*_i(k)),
\end{equation}
\begin{equation}\label{eq:hat2}
E(F_1(k+1)) = E(F_1(k)) + p r - \hat{\beta}_k \ E(F_1(k)) + \hat{\alpha}_k (1-r) \
E(F_1^*(k))
\end{equation}
and
\begin{equation}\label{eq:hat3}
E(F^*_i(k+1)) = E(F^*_i(k)) + \hat{\beta}_k (1-q) i E(F_i(k)) - \hat{\alpha}_k (1- r) i
E(F^*_i(k)).
\end{equation}
\smallskip

In order to obtain an asymptotic solution of (\ref{eq:hat1}), (\ref{eq:hat2}) and
(\ref{eq:hat3}), we require that $E(F_i(k))/k$ and $E(F^*_i(k))/k$ converge to some values
$f_i$ and $f^*_i$, respectively, as $k$ tends to infinity. Assume for the moment that this
is the case, then, provided the convergence is fast enough, $E(F_i(k+1))- E(F_i(k))$ tends
to $f_i$ and  $E(F^*_i(k+1))- E(F^*_i(k))$ tends to $f^*_i$ as $k$ tends to infinity. By
``fast enough'' we mean that $\epsilon_{i,k+1} - \epsilon_{i,k} = o(1/k)$ and
$\epsilon^*_{i,k+1} - \epsilon^*_{i,k} = o(1/k)$ for large $k$, where
\begin{displaymath}
E(F_i(k)) = k (f_i + \epsilon_{i,k}) \quad {\rm and} \quad E(F^*_i(k)) = k (f^*_i +
\epsilon^*_{i,k}).
\end{displaymath}
\medskip

Now, letting
\begin{equation}\label{eq:beta}
\beta = \frac{1-p}{r p + (1-p) q + p (1-r) \psi - (1-p) (1-q) \theta},
\end{equation}
we see that $\beta_k E(F_i(k))$ tends to $\beta f_i$ as $k$ tends to infinity, and letting
\begin{equation}\label{eq:alpha}
\alpha = \frac{p}{(1-p) (1-q) \theta - p (1-r) \psi},
\end{equation}
we see that $\alpha_k E(F^*_i(k))$ tends to $\alpha f^*_i$ as $k$ tends to infinity.

\medskip

So, letting $k$ tend to infinity, (\ref{eq:hat1}), (\ref{eq:hat2}) and (\ref{eq:hat3})
yield, for $i > 1$,
\begin{displaymath}
f_i = \beta \Big(q (i-1) f_{i-1} - i f_i \Big) + \alpha (1-r) i f^*_i,
\end{displaymath}
for $i = 1$,
\begin{displaymath}
f_1 = p r - \beta f_1 + \alpha (1-r) f^*_1,
\end{displaymath}
and for $i \ge 1$,
\begin{displaymath}
f^*_i = \beta (1-q) i f_i - \alpha (1-r) i f^*_i,
\end{displaymath}
whence
\begin{equation}\label{eq:derive-fi-star}
f^*_i = \frac{\beta (1-q) i}{1 + \alpha (1-r) i} \ f_i
\end{equation}
and
\begin{equation}\label{eq:derive-f1}
f_1 = \frac{\varrho p r (\tau + 1)}{(\varrho + 1) (\tau + 1) - (1-q)},
\end{equation}
where $\varrho = 1 / \beta$ and $\tau = 1 / (\alpha (1-r))$. Hence
\begin{displaymath}
f_i = \beta \Big(q (i-1) f_{i-1} - i f_i \Big) + \frac{\alpha \beta (1-r) (1-q) i^2}{1 +
\alpha (1-r) i} \ f_i
\end{displaymath}
and thus
\begin{equation}\label{eq:derive-fi}
f_i =  \frac{q (i-1) (\tau + i)}{(\varrho + i) (\tau + i) - (1-q) i^2} \ f_{i-1}.
\end{equation}

On using (\ref{eq:derive-fi}), repetitively, and (\ref{eq:derive-f1}), the solution to $f_i$ is
given by
\begin{equation}\label{eq:fi}
f_i = \frac{\varrho p r \ \Gamma(i) \Gamma(i + \tau + 1) \Gamma(x + y + 1) \Gamma(x - y +
1)}{\Gamma(\tau + 1) \Gamma(i + x + y + 1) \Gamma(i + x - y + 1)},
\end{equation}
where
\begin{displaymath}
x = \frac{\varrho + \tau}{2 q},
\end{displaymath}
\begin{eqnarray*}
y  \ & = & \frac{\left( (\varrho + \tau)^2 -  4 q \tau \varrho \right)^{1/2}}{2 q},
\end{eqnarray*}
and $\Gamma$ is the gamma function \cite[6.1]{ABRA72}.

\smallskip

Thus for large $i$, on using the asymptotic expansion of the ratio of two gamma functions
\cite[6.1.47]{ABRA72}, we obtain
\begin{equation}\label{eq:fi-approx}
f_i \sim \frac{C}{i^{\frac{\varrho + (1-q) \tau}{q} + 1}},
\end{equation}
where $\sim$ means {\em is asymptotic to} and
\begin{equation}\label{eq:const}
C = \frac{\varrho p r \ \Gamma(x+y+1) \Gamma(x-y+1)}{\Gamma(\tau + 1)}.
\end{equation}
\smallskip

Moreover, it can easily be verified from (\ref{eq:derive-fi-star}) that
\begin{equation}\label{eq:fi-star}
f^*_i = \frac{1-q}{\varrho \left( 1/i + 1/\tau \right)} \ f_i
\end{equation}
and, from (\ref{eq:fi-approx}) and (\ref{eq:fi-star}), it follows that
\begin{displaymath}
f_i + f^*_i \sim \frac{C}{i^{\frac{\varrho + (1-q) \tau}{q} + 1}} \left( 1 + \frac{(1-q)}{\varrho
\left( 1/i + 1/\tau  \right)} \right).
\end{displaymath}
\smallskip


\section{Fitting the Parameters of the Model}
\label{sec:param}

In order to validate the model we use the equations we have derived in Section~\ref{sec:power-law} to fit
the parameters of the model. As a first step we validate the model through stochastic simulation, and
then, in Section~\ref{sec:p2p}, we provide a proof of concept on a real wireless network.

\smallskip

We note that the full set of parameters will, generally, be unknown for real data sets. The
output from each simulation run is the set of unstarred and starred urns, from which we can
infer $balls_k$ and $balls^*_k$, the expected number of balls at stage $k$ in the unstarred and
starred urns, respectively, and $pins_k$ and $pins^*_k$, the expected number of pins in the
unstarred and starred urns, respectively. We are also able to derive approximations for
$balls_k$ and $balls^*_k$, separately, and similarly for pins, based on their definitions in
Section~\ref{sec:urn}.


\smallskip

From the formulation of the model in Section~\ref{sec:urn}, we have
\begin{equation}\label{eq:balls-k}
\frac{balls_k + balls^*_k}{k} \approx pr,
\end{equation}
where the right-hand side of (\ref{eq:balls-k}) is the limiting value of the left-hand
side as $k$ tends to infinity. Similarly, we have,
\begin{equation}\label{eq:pins-k}
\frac{pins_k + pins^*_k}{k} \approx  p r + (1-p) q.
\end{equation}
\smallskip

As a result, we can compute the branching factor, {\em bf}, as
\begin{displaymath}
bf = \frac{pins_k + pins^*_k}{balls_k + balls^*_k},
\end{displaymath}
which eliminates $k$, and derive
\begin{equation}\label{eq:p-bf}
p \approx \frac{1}{r/q \ (bf - 1) + 1}.
\end{equation}
\medskip

The value of the parameter $\varrho$ can be computed from
\begin{equation}\label{eq:rho-k}
\varrho \approx \frac{pins_k}{k (1-p)},
\end{equation}
which follows from (\ref{eq:betak}) and the fact that $\varrho \approx (k \beta_k)^{-1}$.
Similarly, $\tau$ can be computed from
\begin{equation}\label{eq:tau-k}
\tau \approx \frac{pins^*_k}{k p (1-r)},
\end{equation}
which follows from (\ref{eq:alphak}) and the fact that $\tau \approx (k \alpha_k
(1-r))^{-1}$. Moreover, the value of the constant $C$ can be derived from
(\ref{eq:const}), given $p, q, r, \varrho$ and $\tau$.

\medskip

To fit the parameters we can now numerically minimise the least squares of
\begin{equation}\label{eq:sumsq}
\sum_i^m \mid urn_i \mid - \ C k f_i,
\end{equation}
where $k$ is the number of steps in the simulation, $\mid urn_i \mid$ denotes the number of
balls in $urn_i$, $m$ denotes the number of urns over which the minimisation takes place and
$f_i$ is given by (\ref{eq:fi}), in order to estimate one or more of the parameters given
knowledge of the others. (For a justification of choosing $m$ to be the first gap in the urn
set, i.e. such that from $i=1$ to $m$ $urn_i$ is non-empty and $urn_{m+1}$ is empty, see
\cite{FENN02}.)

\medskip

We note that we have chosen to do a direct numerical minimisation rather than use a regression tool
on the log-log transformation of the urn data and try to fit a power-law distribution, since fitting
power-law distributions is problematic \cite{GOLD04}. Moreover, the $f_i$'s in our model obey only
asymptotically a power-law distribution and therefore we preferred to fit the ``correct''
distribution with the ratio of gamma functions, as given in (\ref{eq:fi}).

\smallskip

To validate the simulation we fixed the input parameters $p, q, r$ and $k$ and simulated the
model in Matlab as described at the beginning of Section~\ref{sec:urn}. We fixed $q = 0.9$ and
the number of simulation steps to be $k = 10^6$, and varied $p$ and $r$.

\smallskip

We first set $p=0.1$ and $r=0.5$. A typical output of the simulation run produced $balls_k =
10762$, $balls_k^* = 39200$, $pins_k = 77452$ and $pins_k^*  = 39200$. The left-hand side of
(\ref{eq:balls-k}) gives an approximation of $pr$ as $0.05$, while its right-hand side gives
the same value. Correspondingly, the left-hand side of (\ref{eq:pins-k}) gives an approximation
of $pr + (1-p) q$ as $0.8602$, while its right-hand side gives the value $0.86$. Finally, the
left-hand side of (\ref{eq:p-bf}) is just $p$, while its right-hand side gives the approximated
value $p = 0.0999$.

\smallskip

Computing an estimate of $\varrho$ from (\ref{eq:rho-k}) gives $0.0861$, while an
estimation of $\tau$ from (\ref{eq:tau-k}) gives $15.6541$. In order to estimate $\varrho$
and $\tau$ from the urn data, we first fixed all the parameters in (\ref{eq:fi}) apart
from $C$ of (\ref{eq:const}), which we estimated, using (\ref{eq:sumsq}), to be $C =
651950$. We then fixed $C$, given in (\ref{eq:const}), and numerically estimated $\varrho$
and $\tau$ in turn obtaining $\varrho = 0.0865$ and $\tau = 15.6541$.

\smallskip

We next set $p=0.2$ and $r=0.7$. A typical simulation run produced $balls_k = 122179$, $balls_k^* =
18997$, $pins_k = 658273$ and $pins_k^*  = 201521$. The left-hand side of (\ref{eq:balls-k}) gives
an approximation of $pr$ as $0.1406$, while its right-hand side gives the value $pr = 0.14$. The
left-hand side of (\ref{eq:pins-k}) gives as approximation of $pr + (1-p) q$ as $0.8594$, while its
right-hand side gives the value $0.86$. Finally, the left-hand side of (\ref{eq:p-bf}) is just $p$,
while its right-hand side gives the approximated value $p = 0.2009$.

\smallskip

Computing an estimate of $\varrho$ from (\ref{eq:rho-k}) gives $0.8228$, while an estimate of $\tau$
from (\ref{eq:tau-k}) gives $3.3587$. In order to estimate $\varrho$ and $\tau$ from the urn data,
we first fixed all the parameters in (\ref{eq:fi}) apart from $C$ of (\ref{eq:const}), which we
estimated, using (\ref{eq:sumsq}), to be $C = 15742$. We then fixed $C$ in (\ref{eq:fi}) and
numerically estimated $\varrho$ and $\tau$ in turn obtaining $\varrho = 0.7983$ and $\tau = 3.35$.
Additional runs of the simulation produced similar results in terms of their accuracy. We note that
we limited $m$ in (\ref{eq:sumsq}) so that its maximum value be 90, due to numerical overflow of the
product of gamma functions for larger values of $m$.

\smallskip

The simulations demonstrate that, given that the data is consistent with the urn transfer model we
have defined in Section~\ref{sec:urn}, numerical optimisation can be used to accurately estimate the
parameters of the model.

\section{Real Social Networks}
\label{sec:p2p}

As a proof of concept we made use of a public log containing traces of the activity of users
within a campus-wide WLAN network recorded by the Crawdad project
(\url{http://crawdad.cs.dartmouth.edu}) at the Center for Mobile Computing at Dartmouth College
\cite{KOTZ05a}. The data set we elected to work with was collected during 2001-2003 using the
syslog system event logging facility available on the wireless access points. Each access point
was configured so as to transmit a message logged at one of two dedicated servers maintained by
the project, every time a client card authenticated, associated, reassociated, disassociated or
deauthenticated with the access point. In total, approximately $13.5$ million events have been
recorded during this period.

\smallskip

In the syslog records, client cards are identified by their MAC address. It should be noted that
there is no one-to-one relationship between card addresses, devices and users, as in some cases one
card may have been used with more than one device and one device may have been using more than one
card. Moreover, a user may be using more than one device. Mobility traces were computed from the raw
syslog messages for each device. A special access point name signifies that a card is not connected
to the wireless network. This condition was determined by the syslog message ``Disauthentication''
from the last associated access point with reason field ``Inactivity''. Such messages are commonly
generated when the card is inactive for 30 minutes. For simplicity, from now on, we will refer to a
client card as a user.

\smallskip

\begin{figure}[ht]
\centerline{\includegraphics[]{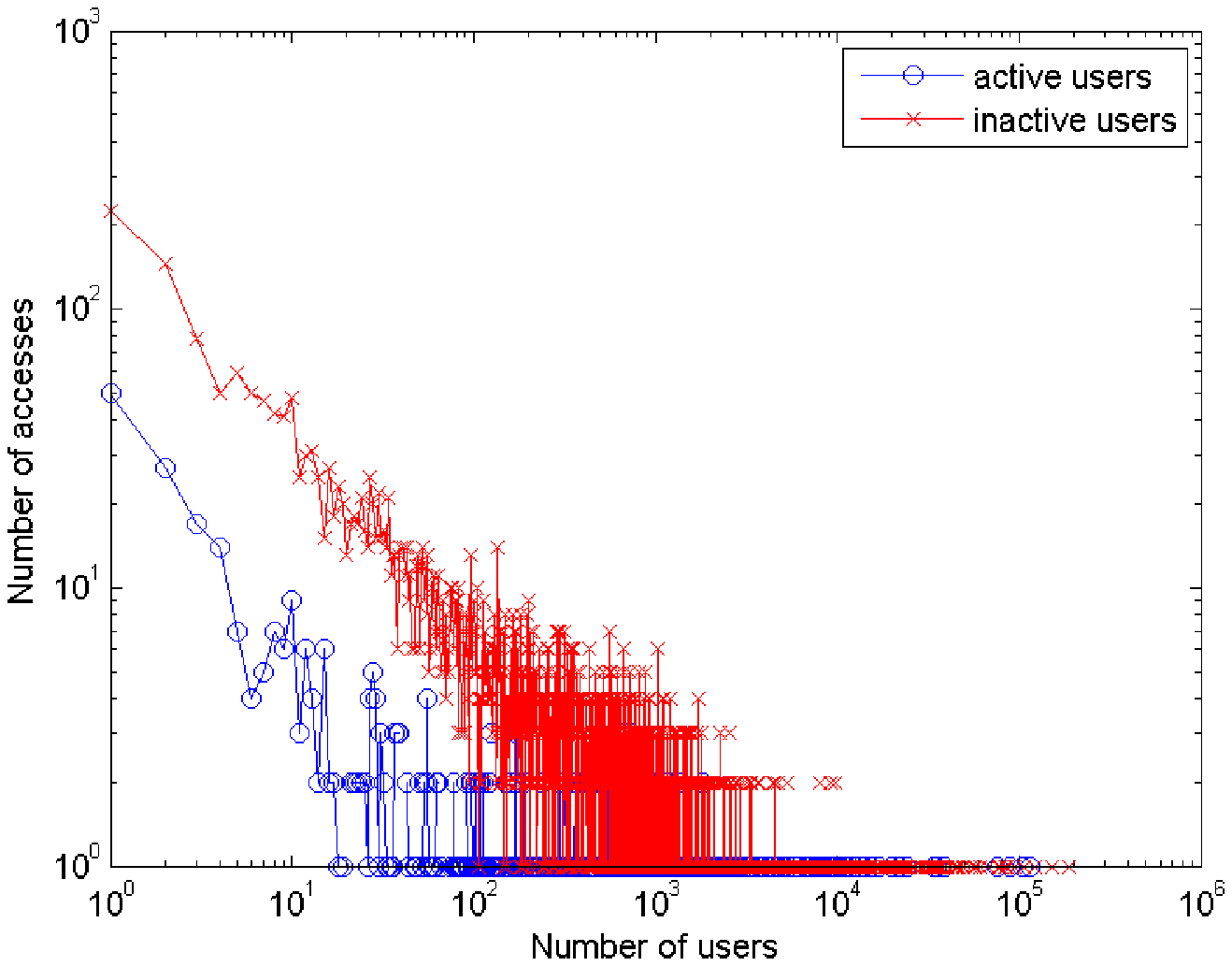}} \caption{\label{fig:users} Log-log plot of
wireless users' activity}
\end{figure}

In Figure~\ref{fig:users} we show the log-log plot of the number of accesses of the active and
inactive users at the end of the trace period. From the figure we may conjecture an asymptotic
power-law distribution, but as can be seen the tails are very fuzzy and therefore regression or
maximum likelihood methods are unlikely to succeed \cite{GOLD04}. For this reason, as mentioned in
Section~\ref{sec:param}, we preferred to estimate the parameters of the model numerically via least
squares minimisation.

\medskip

Our model is fully specified by the four input parameters $p, q, r$ and $k$, as described in
Section~\ref{sec:urn}. Of particular interest are the following probabilities:

\renewcommand{\labelenumi}{(\arabic{enumi})}
\begin{enumerate}
\item $p r$, which is the rate at which new users join the network and attain their first
wireless connection.

\item $p (1-r)$, which is the rate at which inactive users become active again.

\item $(1-p) q$, which is the rate at which active users attain a new wireless connection
without first disconnecting from the network.

\item $(1-p) (1-q)$, which is that rate at which active users become inactive.

\item $k$, which can be viewed as the life of the network, assuming that the evolution takes
place in discrete time steps, where at each time a single change occurs in the network according to
the urn transfer model described in Section~\ref{sec:urn}.

\end{enumerate}
\smallskip

We processed the Dartmouth data set so that it contains pairs of users and their activity, where
each user is identified by a client card and an activity corresponds to (1), (2), (3) or (4) above.
We then estimated the probabilities $p, r$ and $q$ from the data, taking $k$ to be the number of
pairs processed. From this we obtained, $p = 0.0994$, $r = 0.0046$, $q = 0.8897$ and $k = 13559701$.

\smallskip

Next we estimated $\varrho$ from (\ref{eq:rho-k}) and $\tau$ from (\ref{eq:tau-k}), obtaining
$\varrho = 0.1244$ and $\tau = 6.9704$. Using (\ref{eq:fi-approx}) and (\ref{eq:const}) we estimated
the exponent of the asymptotic power-law distribution as
\begin{displaymath}
\frac{\varrho + (1-q) \tau}{q} + 1 = 2.0040.
\end{displaymath}
\smallskip

As a validation of the model we populated the unstarred and starred urns according to the activity
pairs from the processes data set. Then, using the methodology described in Section~\ref{sec:param},
we numerically minimised the least squares of the sum over $i$ of the differences between the number
of balls in $urn_i$, respectively $urn^*_i$, and the predicted number of balls according to
(\ref{eq:fi}), and respectively (\ref{eq:fi-star}), in accordance to (\ref{eq:sumsq}). The fitted
parameters we obtained from the unstarred urns using (\ref{eq:fi}), were: $q = 0.8901$, $\varrho =
0.1101$ and $\tau = 6.9648$, obtaining $(\varrho + (1-q) \tau) / q + 1 = 1.9836$. The corresponding
set of fitted parameters obtained from the starred urns using (\ref{eq:fi-star}), were: $q =
0.8898$, $\varrho = 0.1385$ and $\tau = 6.9473$, obtaining $(\varrho + (1-q) \tau) / q + 1 =
2.0161$. As can be seen the fitted parameters are consistent with the ones we have mined from the
original data set.

\smallskip

As a further validation of the model we ran a simulation implemented in Matlab according to the
description of the stochastic process in Section~\ref{sec:urn}, with the parameters $k =
13559701$, $p = 0.0994$, $r = 0.0046$ and $q = 0.8897$ as mined from the data set. We note that
\begin{displaymath}
p = 0.0994 > (1-p) (1-q) = 0.0993
\end{displaymath}
and
\begin{displaymath}
(1-p) (1-q) = 0.0993 > p (1-r) = 0.0989
\end{displaymath}
as required in the specification of the stochastic process in Section~\ref{sec:urn}. So, for
the probability of termination, with either all starred or unstarred urns being empty, to be
less than $0.1$ we should set the initial number of balls in $urn_1$ to be $\delta = 3600$, and
the initial number of balls in $urn^*_1$ to be $\delta^* = 600$. We verified this by running a
simplified version of the simulation, which only accounts for the total number of balls in
starred and unstarred urns. Out of ten simplified simulation runs with the above input
parameters none terminated with all the unstarred or starred urns being empty.

\smallskip

We decided in our simulation to ignore the problem of empty urns, the justification being that
having empty urns at some stage of the stochastic process does not have much effect on the
exponent of the asymptotic power-law distribution, since by (\ref{eq:rho-k}) and
(\ref{eq:tau-k}) the exponent given in (\ref{eq:fi-approx}) is approximately proportional to
$pins_k + pins^*_k$, and by (\ref{eq:pins-k}) the total number of pins depends only on the
input parameters through independent random variables.

\smallskip

From $pins_k$ and $pins^*_k$ output from the simulation we computed $\varrho = 0.1054$ from
(\ref{eq:rho-k}), $\tau = 7.1442$ from (\ref{eq:tau-k}), and finally the exponent of the asymptotic
power-law distribution was computed as $(\varrho + (1-q) \tau) / q + 1 = 2.0042$. As can be seen,
the output from the simulation is consistent with the parameters mined from the data; a second
simulation with the same input parameters produced similar results.

\smallskip

Overall, on the evidence from the computational results , the urn transfer model described in
Section~\ref{sec:urn}, is a viable model for a real social network, specifically for the access
patterns of users within the Dartmouth wireless network.

\section{Concluding Remarks}
\label{sec:concluding}

We have presented an extension of Simon's classical stochastic process where each actor can be
either in an active or an inactive state. Actors, chosen by preferential attachment may attain a new
relation, become inactive or later become active again. The system is closed in the sense that once
an actor enters the system he remains within the system. We have shown in (\ref{eq:fi-approx}) and
(\ref{eq:fi-star}) that, asymptotically, the number of active and inactive actors having prescribed
number of relations is a power-law distribution. As a proof of concept we validated the model on a
real data set of wireless accesses over a lengthy period of time. The validation made use of
numerical optimisation rather than using standard regression tools, due to the known difficulty of
detecting asymptotic power-law distributions in data.

\medskip

The stochastic model we have presented is relevant to social networks where users may be active or
inactive at different times. Two such real-world networks are wireless networks and peer-to-peer
networks, although it remains to validate our model on a real peer-to-peer data set. In fact, our
model could also be used to model user activity in an e-commerce portal or an online forum, where
registration is required.

\newcommand{\etalchar}[1]{$^{#1}$}

\end{document}